\documentstyle[12pt,epsfig]{article}
\def\ifig#1#2#3#4{\begin{figure}[hbtp]
 \begin{center}\leavevmode\epsfig{file=#1,height=#2}
 \caption[#3]{#4}\end{center}\end{figure}}
\def\figlr#1#2#3#4#5{% left-right
  \begin{figure}[hbtp]\begin{center}\leavevmode
  \hfill\epsfig{file=#1,height=#3}\hfill\epsfig{file=#2,height=#3}\hfill\hfill
  \caption[#4]{#5}\end{center}\end{figure}}
\def\figlcr#1#2#3#4#5#6{
 \begin{figure}[hbtp]\begin{center}\leavevmode
 \hfill\epsfig{file=#1,height=#4}\hfill\epsfig{file=#2,height=#4}\hfill
 \epsfig{file=#3,height=#4}\hfill\hfill
 \caption[#5]{#6}\end{center}\end{figure}}

\title{ Decay of hot, rotating, compound nuclei}

\author{             Krzysztof Pomorski
    \thanks{On leave on absence from University M.C.S. in Lublin},
              Klaus Dietrich, Wojciech Przystupa
\thanks{Permanent address: Agriculture University, Lublin, Poland}\\
      Technische Universit\"at M\"unchen, Garching, Germany\\
                 Johann Bartel and Jean Richert\\
      Laboratoire de Physique Th\'eorique, Universit\'e Louis Pasteur\\
                      Strasbourg, France}

\date{}

\begin{document}

\maketitle

\begin{abstract}
The fusion process and the competition between fission and $n$, $p$ and
$\alpha$--particle emission is studied.  The calculations are performed 
for nuclei at excitation energies from 80~MeV up to about 300~MeV.  
The nuclear fission is described by a~Langevin equation coupled to 
the Master equation for particle evaporation. A significant influence 
of the initial spin distribution on the prescission particles multiplicities 
is found. 
\end{abstract}

\section{Introduction}

The present paper is a continuation of our previous work 
\cite{Str91,Pom96}.  We are studying systems with excitation energies 
of the order of (80--300) MeV.  We assume that the
particle emission is described by the Wei{\ss}kopf theory \cite{Wei39},
and that the nuclear fission is a~transport process \cite{Kra40}. The
emission of photons is neglected because at these excitation energies
the particle evaporation is expected to be dominant.
 
Grang\'e and Weidenm\"uller \cite{Gra79} were the first point out the 
importance 
of the non--statistical aspects of the fission process. Our model and also 
the ones of Fr\"obrich, Abe and Carjan \cite{Fro92,Til92,Abe90} are based 
on their work. In the last years one can observe the increasing amount of 
experimental studies of the evaporation of light particles from excited 
nuclei and of the concomitant decay by fission \cite{Hil92}. Results 
of our calculations are presented and compared with experimental work, 
especially that one of Ref. \cite{Gon90}. The present research is closly 
correlated with the measurements performed by the DEMON group in Strasbourg 
\cite{Han97}.

In our model we assume that the fission process is described by
a Langevin equation which is dynamically coupled with a Master type
equation for the light particles evaporation. We take into account the
dependence of the evaporation probabilities on the deformation of 
nucleus, on its excitation and  collective rotation. We assume that 
the transmission coefficient depends on deformation and on
collective rotation of nucleus. The collective potential
of the fissioning nucleus is evaluated within the model of a deformed, 
hot and rotating liquid drop \cite{Bar95}. The effective one--dimensional path
to fission is chosen in a three--dimensional deformation space.
The collective inertia is obtained in the irrotational flow model
and the wall formula \cite{Blo86} is used to evaluate the strength of
the friction forces. The Einstein relation between the friction and diffusion
parameters is assumed to hold.

\section{Results} 

We study the decay of the compound nuclei $^{160}$Yb and $^{126}$Ba at various 
excitation energies ranging from 80 MeV to 300 MeV.  It is 
of special interest to investigate the influence of deformation and fast
rotation on the emission of $n$, $p$ and $\alpha$--particles from excited 
states of these nuclei.  
\subsection{Decay of $^{160}$Yb}

We have selected this nucleus because a~careful experimental 
investigation of its decays is available \cite{Gon90} 
and because it has a large ground state deformation.
In Fig. 1, the deformation dependent emission width for
$n$, $p$ and $\alpha$ is shown for two different isotopes of both 
$_{64}$Gd and $_{70}$Yb. These 4 nuclei were chosen in order to illustrate 
how the deformation dependence of the emission widths for $n$,$p$, and 
$\alpha$ varies with the neutron and proton number of the emitting nucleus. 
\figlcr{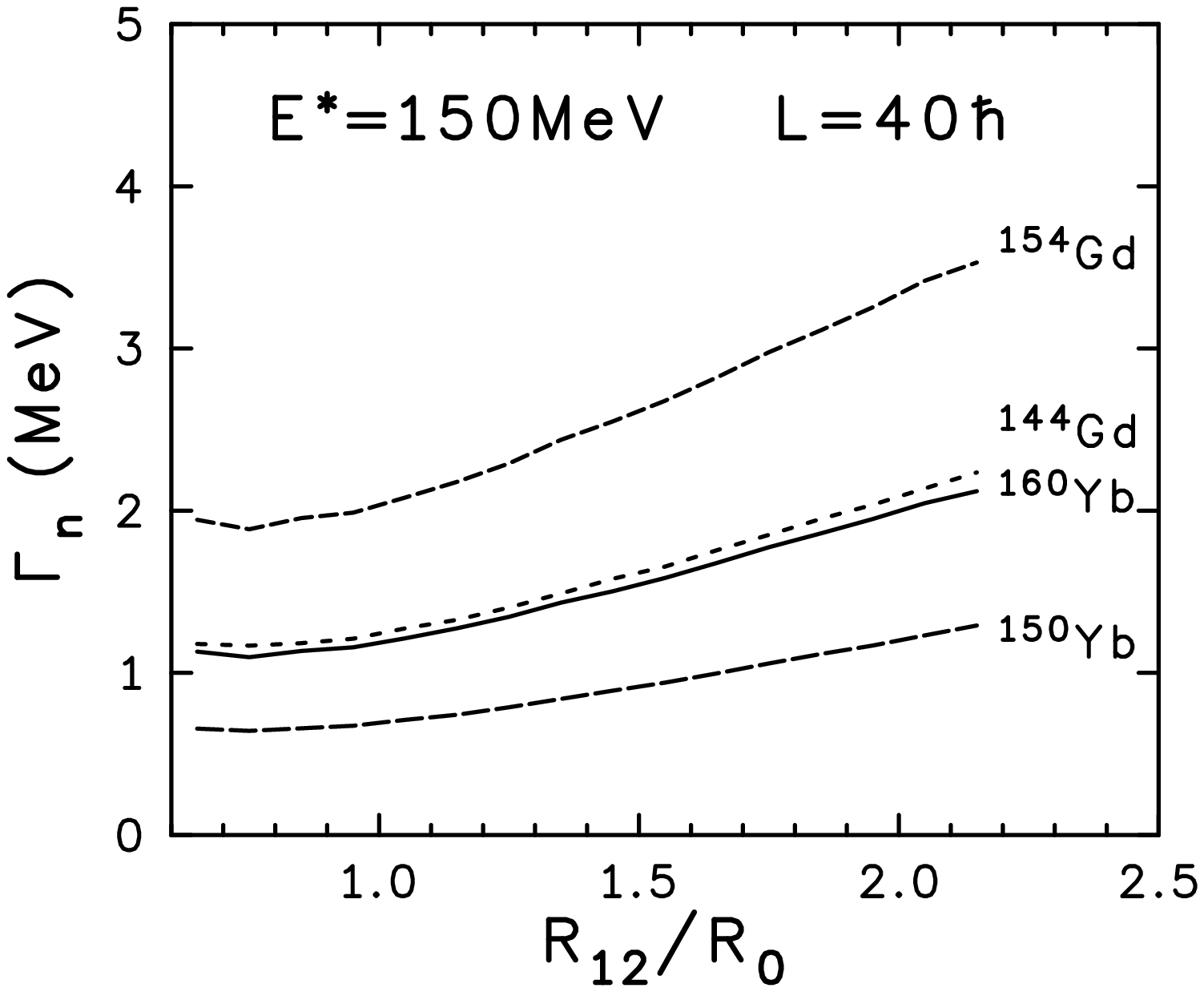}{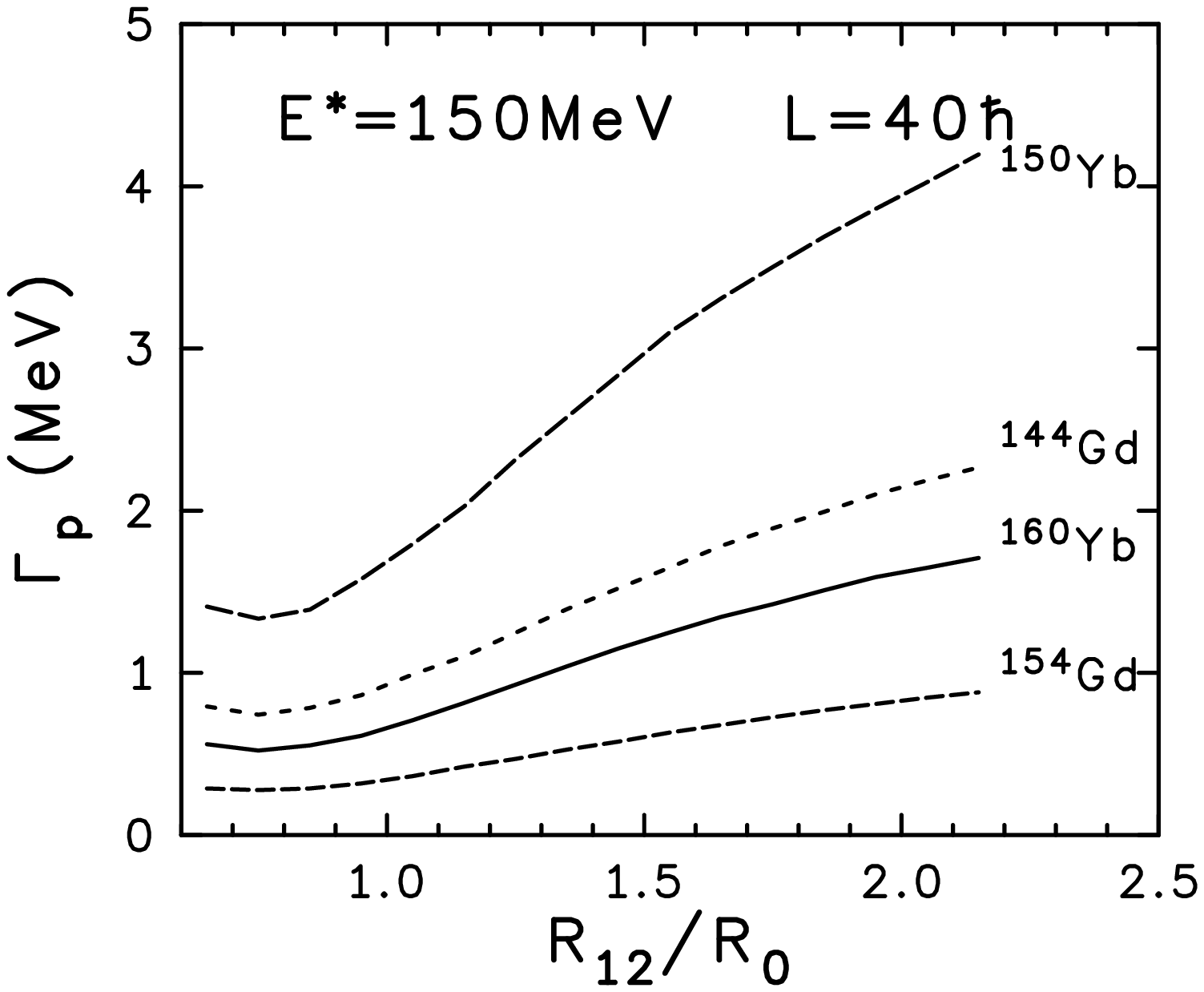}{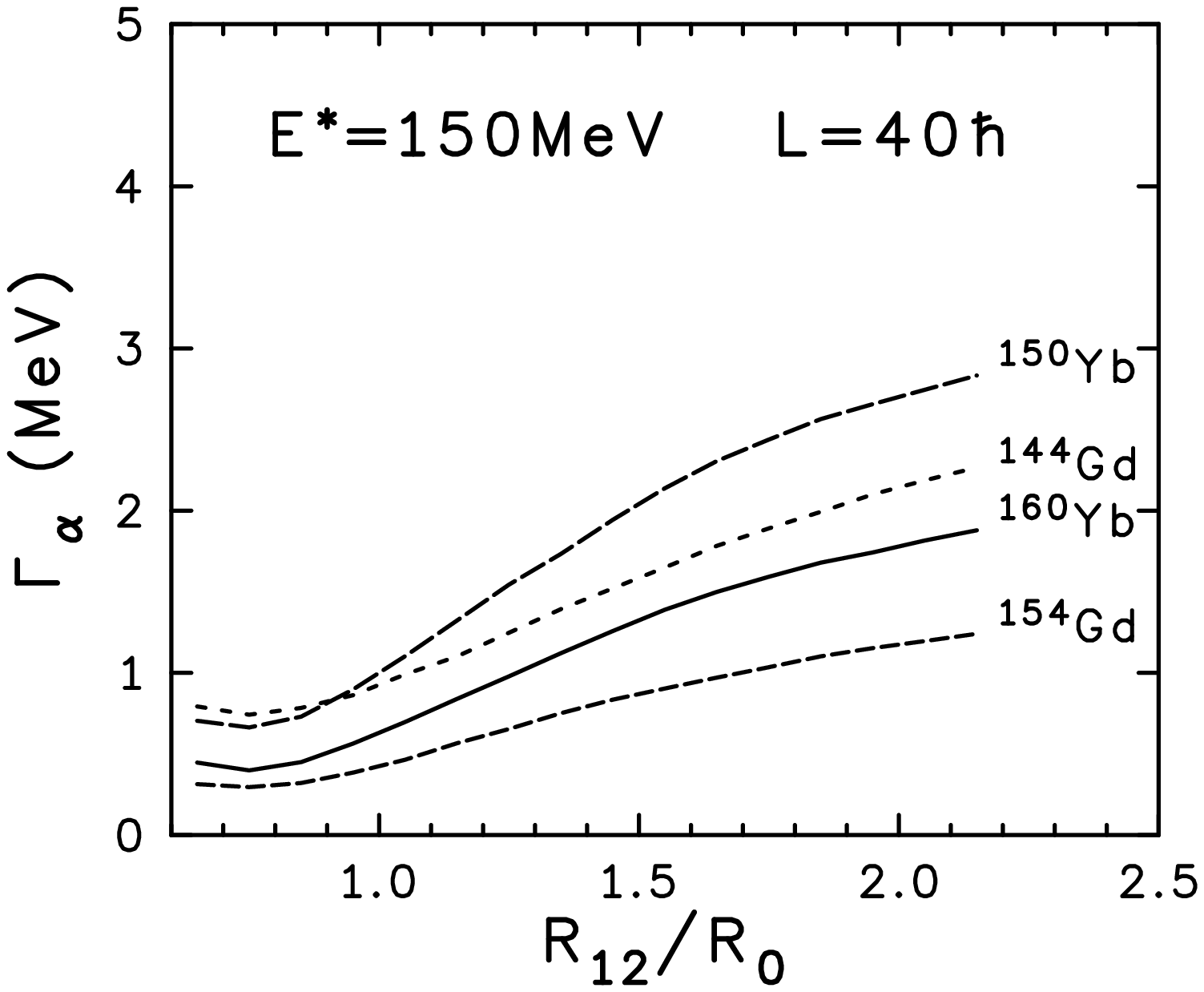}{36mm}{A}{}
%1
All the emission rates grow as a~function of increasing
deformation. This trend can be easily understood as for increasing
deformation the transmission occurs through a~larger surface. The
effect has already been observed for all three types of particles \cite{Bla80}.
We notice, however, that the emission width for $\alpha$--particles
increases more steeply than the one for $n$ and $p$ for lower
excitation energies.  This is due to the fact that, as the nucleus is
elongated, the barrier height for charged particles is reduced in the
section of the surface which is farther away from the nuclear center
and increased in the section which is closer to the center. Consequently, 
the emission rate for charged particles
increases faster than the one for neutrons.

The dependence of the number of decays of a given type on the
time which elapses starting from the formation of the compound
nucleus is unfortunately not measurable. Nevertheless, it is interesting 
to study this dependence theoretically in order to illustrate the time 
scale of this process. In Fig. 2 we show the number of fission
events as a function of the time t on a logarithmic scale. 
\figlr{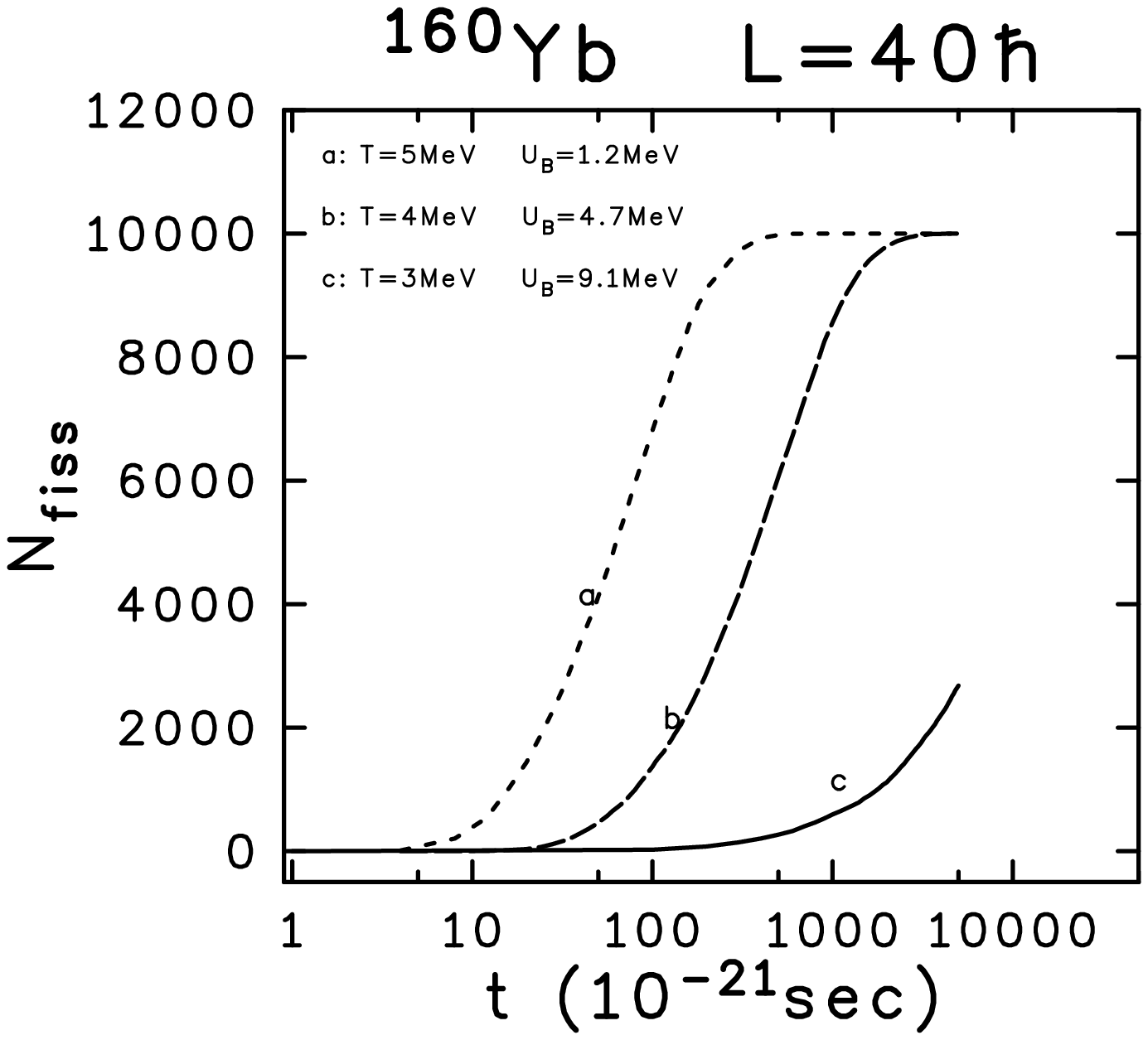}{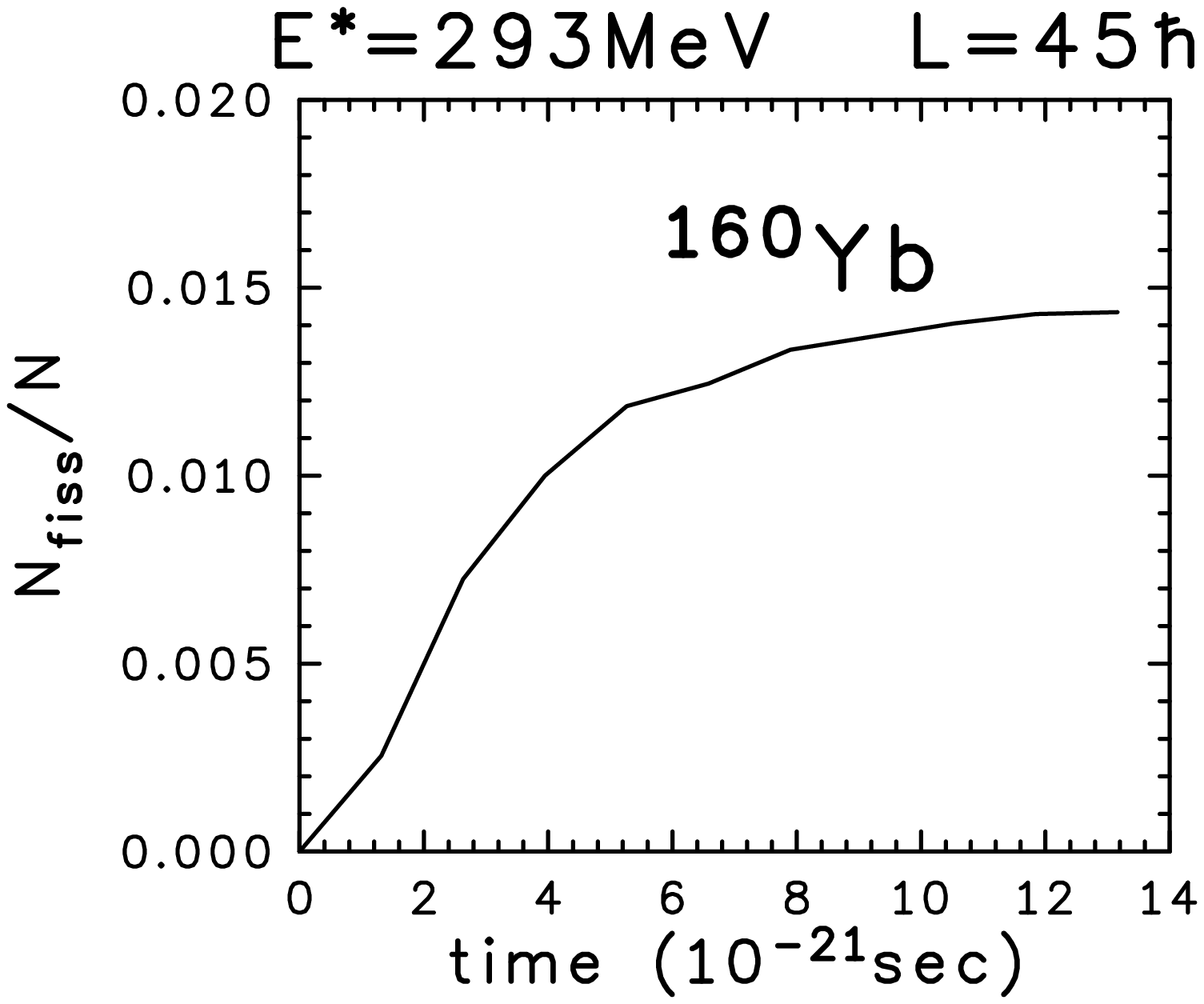}{50mm}{2}{} 
%2
This result was 
obtained with the light-particle evaporation channels turned off. The initial
compound nucleus is $^{160}_{70}$Yb with an initial angular
momentum $L = 40~\hbar$. The three curves on the l.h.s. of Fig. 2 correspond 
to 3 different initial temperatures resulting in 3 different
initial fission barrier heights $U_B$. It is seen that the transient
time increases with decreasing excitation energy, as one
expects. Please note that the transient
time interval is seen to be totally different from the one in
the Kramers regime. The Kramers limit is valid in the cases when the fission
barrier is much higher than the temperature of the fissioning nucleus.
On the r.h.s. of Fig. 2 the fraction ${N_{\rm fiss}/N}$ of nuclei 
undergoing fission is shown for the initial compound nucleus 
$^{160}_{70}$Yb as a function of time. 
Now, contrary to the results presented in the l.h.s. of Fig. 2, the
emission of light particles is taken into account.  This implies that
the fission barrier rises as a function of time, since, at each particle
emission act, the excitation energy and (on the average) the angular
momentum of the emitting nucleus decreases. Consequently, the time
scale of the fission process is stretched.  The initial excitation
energy of the nucleus is $E^{\star} = 293$~MeV ($T \approx 5$~MeV) and 
the initial angular momentum $L = 45~\hbar$ is assumed.

The multiplicity of  prefission neutrons, protons and $\alpha$--particles 
is plotted in Fig. 3 as a function of the initial angular momentum 
$L$. 
\figlr{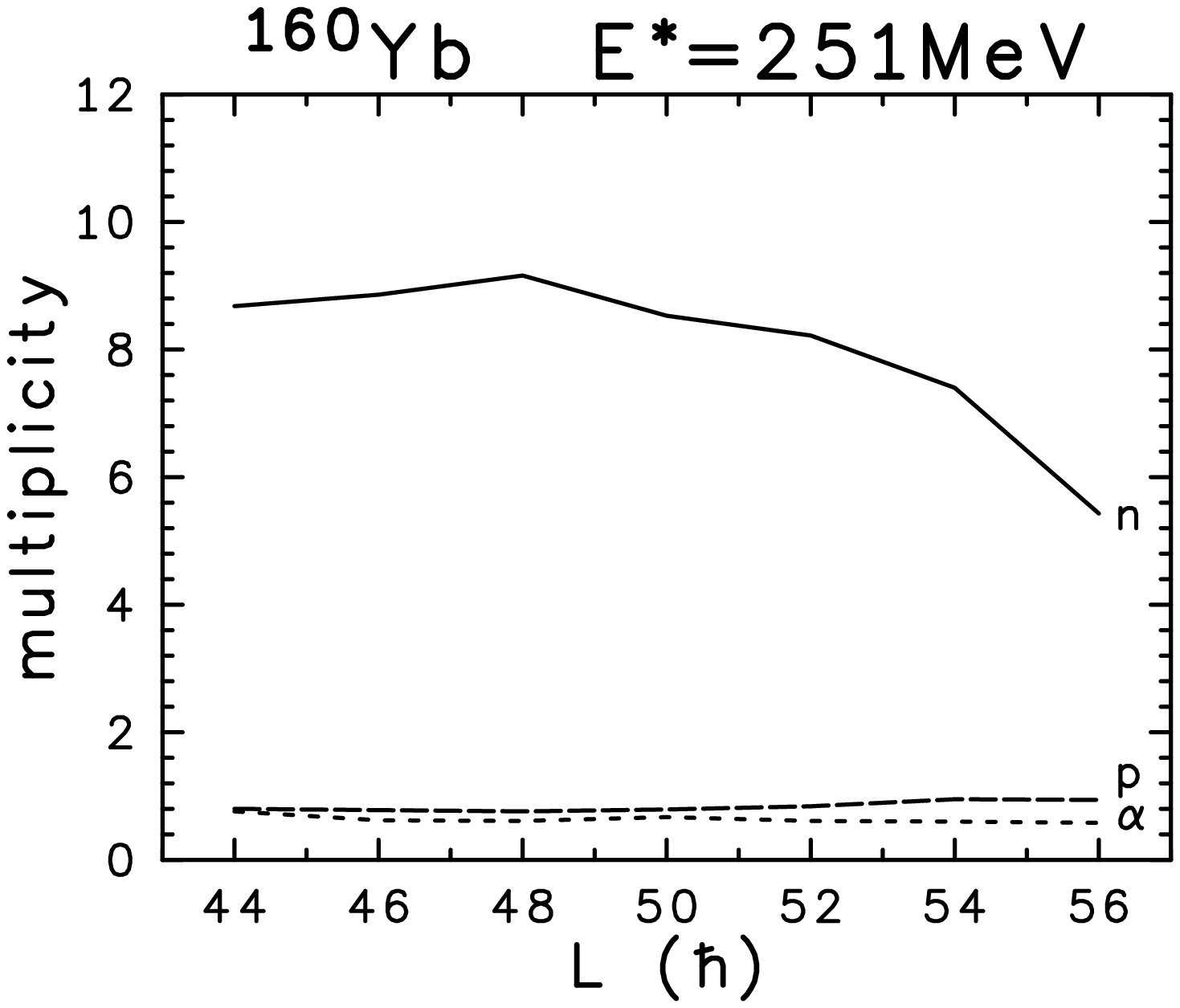}{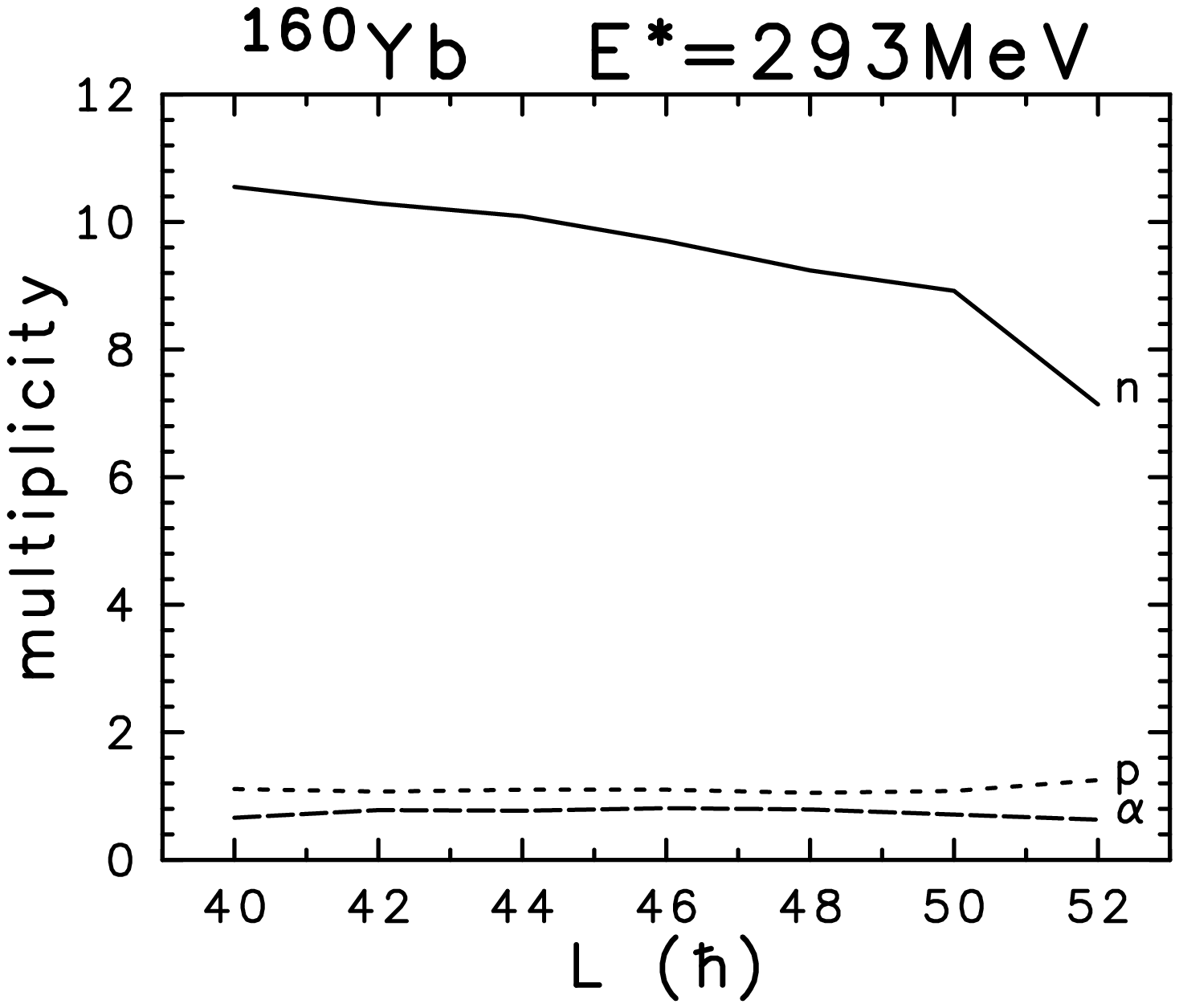}{50mm}{6}{}%
%3
The initial compound nucleus is $^{160}_{70}$Yb with 
the initial excitation energy E$^*$=251 MeV (l.h.s.) and 293 MeV (r.h.s.).
As to be seen in Fig. 3, the neutron multiplicity decreases 
significantly with growing $L$ while that for protons and $\alpha$--particles 
is much less affected. This is due to the fact that for large angular momenta 
the fission barriers $U_B$ become small  and it takes a shorter time to reach 
the scission configuration. Consequently less neutrons are emitted on the 
average when the fission barrier is small. $\alpha$--particles and protons are
mostly emitted in the initial stage when the excitation energy of the nucleus 
is large, so that their multiplicities depend less strongly on the initial 
$L$.

This result indicates that a more precise knowledge of the initial spin
distribution in compound nuclei is necessary for a meaningful comparison 
with experiment. For the case of $^{160}$Yb the reduction of the fission 
barrier ($U_B$) by  fast rotation has the consequence that only the largest 
$L$ values contribute to the measurable values of the multiplicities as one 
can see in Fig. 4.  
\figlr{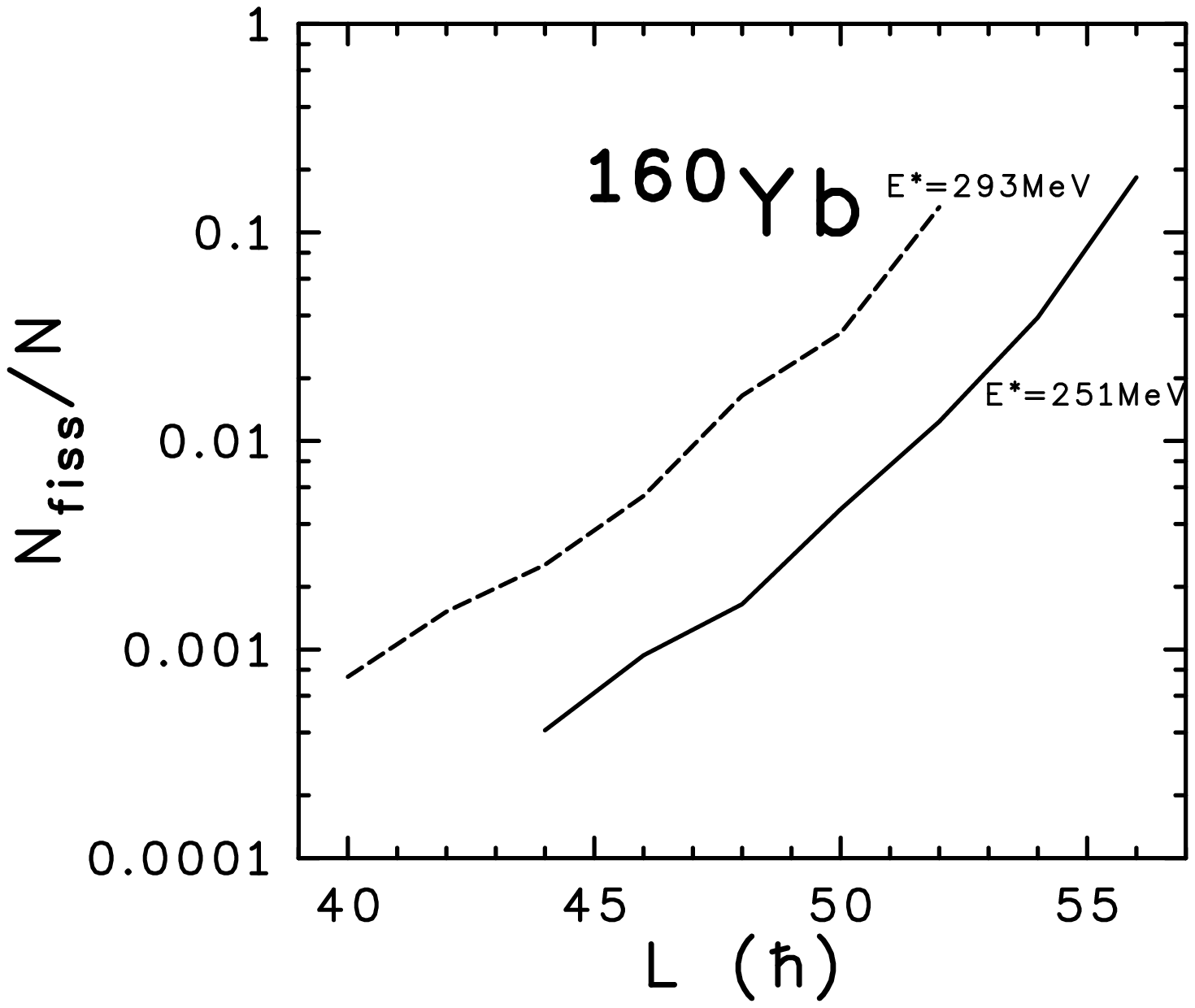}{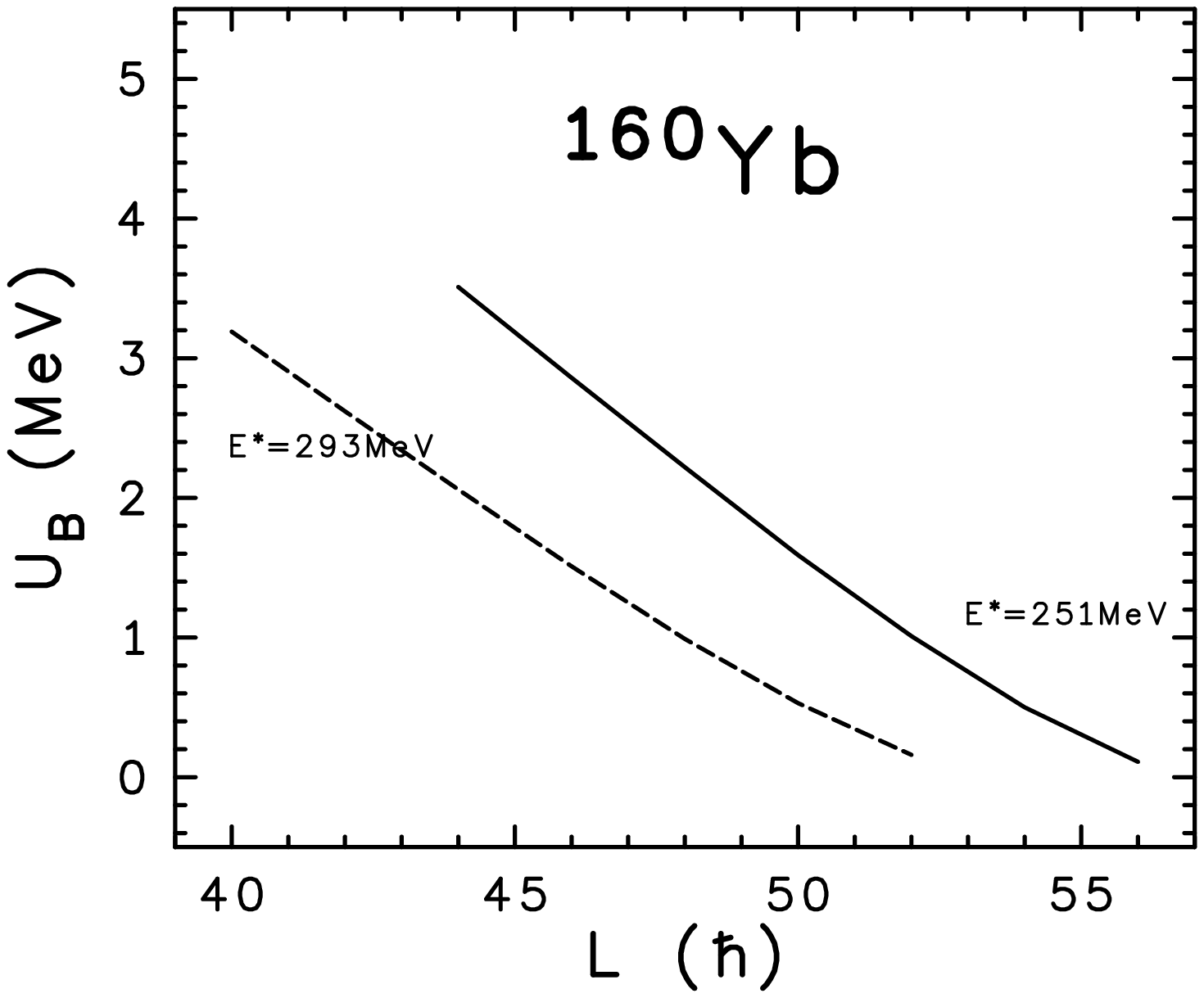}{50mm}{8}{}
%4
The theoretical estimates of the prefission particle multiplicities,
averaged over all angular momenta of compound nucleus, are compared in 
Table 1 with the experimental data taken from Ref. \cite{Gon90}.

\begin{table}[h] 
\begin{center} 
\begin{tabular}{|c|c|c|c|c|} 
\hline 
 & \multicolumn{2}{c|}{$E^*$=251 MeV}   &  \multicolumn{2}{c|}{$E^*$=293 MeV}\\ 
\hline 
 
 $\nu$   &   model &  exp.  &   model  &  exp. \\ 
\hline 
   n     &    5.98 & 6.10$\pm$ 1.5  &    7.80  & 8.50$\pm$ 1.6  \\ 
\hline 
   p     &    0.94 & 0.51$\pm$ 0.07   &    1.19  & 0.70$\pm$ 0.08  \\ 
\hline 
$\alpha$ &    0.58 & 0.48$\pm$ 0.07   &    0.66  & 0.75$\pm$ 0.08  \\ 
\hline 
\end{tabular} 
\end{center} 
\end{table} 

All parameters of the model are given in Ref. \cite{Bar95}.  We only
have to choose a preformation factor $f_\alpha = 0.2$ for
reproducing the experimental number of $\alpha$--particles. This goes into
the right direction since our calculations show that $\alpha$--particle
emission is strongly enhanced by rotation and deformation effects.  One
of the most important further improvements of the theory will be to
evaluate this preformation factor within the temperature--dependent
Thomas--Fermi approximation which underlies our theory \cite{Die96}. At
low temperature, the Thomas-Fermi approximation is expected to yield
too low values of the preformation factor. Additional effects like the
pairing correlations will then increase the value of the preformation
factor at low excitation energies (T $\leq$ 1 MeV).

In the l.h.s. of Fig. 5, we present the energy spectra
of neutrons (n), protons (p) and $\alpha$--particles emitted by fissioning 
nuclei (solid lines) and by the fission residua (dashed lines). 
\figlr{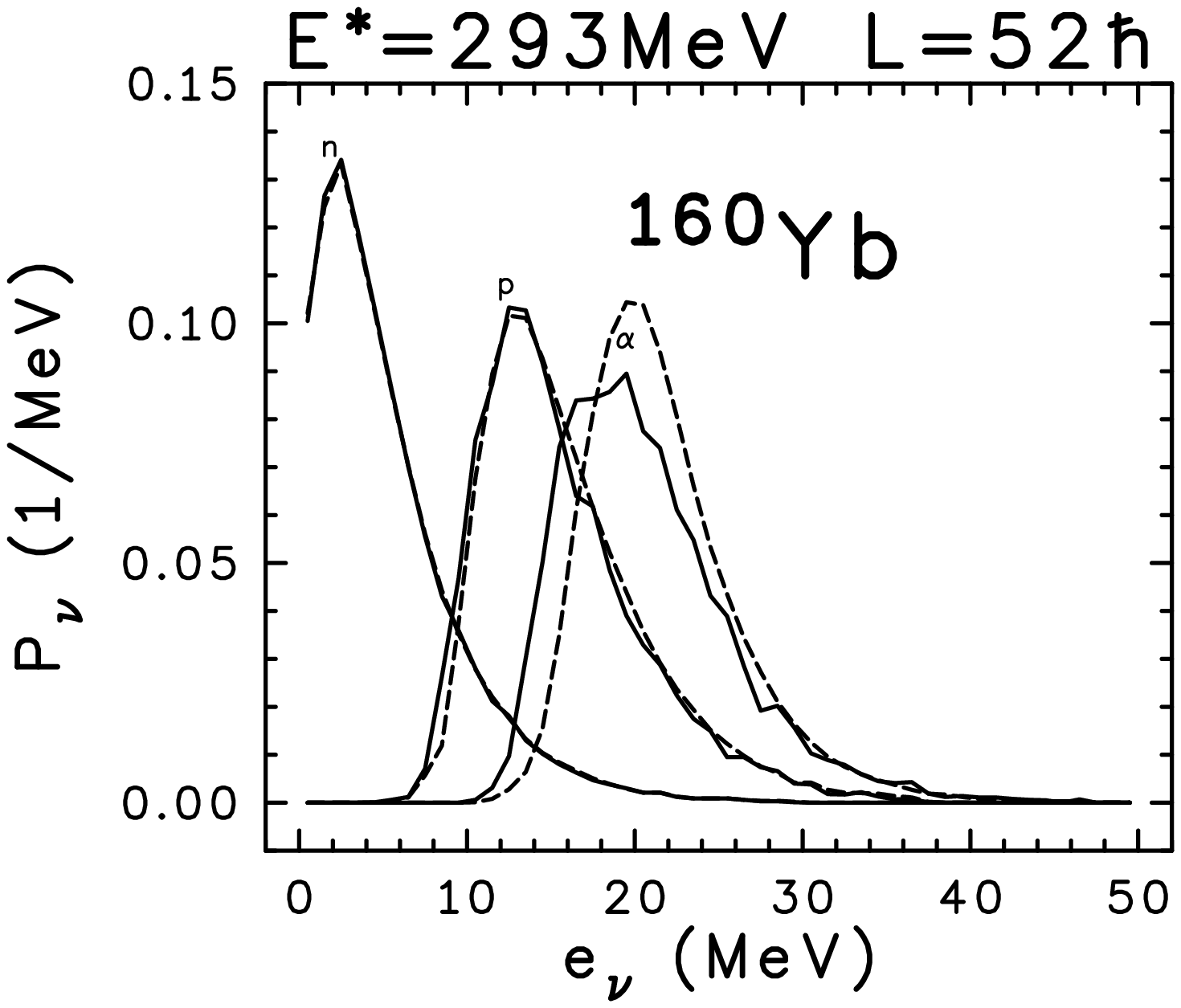}{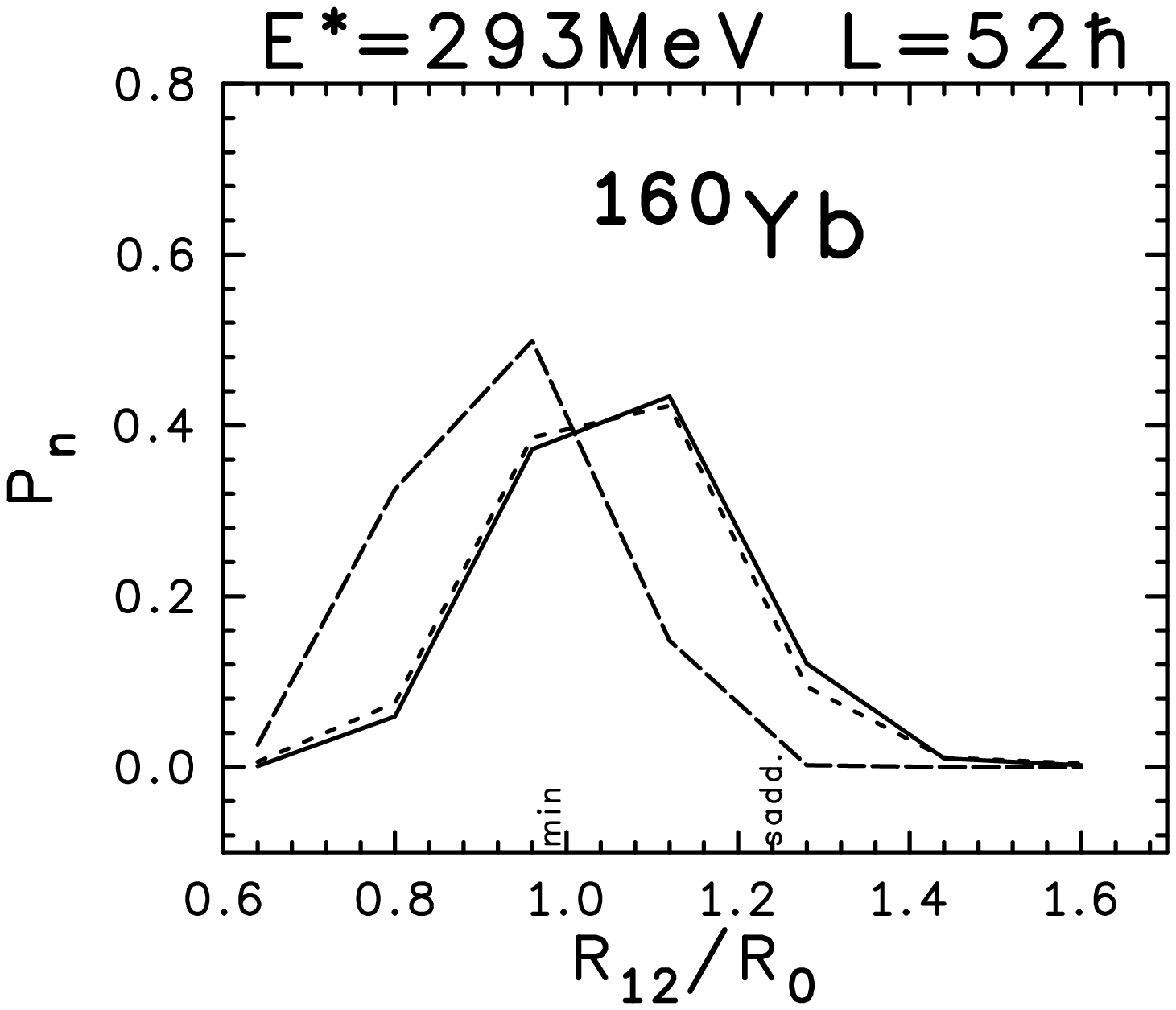}{50mm}{9}{}
%5

The spectral distribution for $\alpha$--particles emitted in
coincidence with fission is shifted by about 2 MeV towards smaller
energies as compared to the distribution obtained when measured in
anti-coincidence with fission, while for neutrons the both
distributions are very close to each other. This is due to the fact that
charged particles are preferentially emitted from the pole tips around
the long half axis. The larger deformation then implies a smaller gain
of kinetic energy from the repulsive Coulomb field.

In the r.h.s. of Fig. 5,  the normalized yield for neutron emission is
shown as a function of the nuclear deformation in the fission and
evaporation channels. One can notice that for nuclei which undergo
fission, the emission of neutrons takes place, on the average, at a
larger deformation than for the nuclei which end up as evaporation
residues. This is due to the fact that the distribution of the
fissioning nuclei moves towards the saddle point, i.e. into a region of
increasing deformation. This effect should
give rise to an experimentally observed anisotropy in the angular
distribution of prefission particles different from the one observed in
the angular distribution of particles emitted by the evaporation
residua: We expect that less neutrons will be emitted in the
reaction plane than in the direction perpendicular to this plane. The
anisotropy in the angular distribution of the charged particles will be
probably smaller than that in the neutron case. This is due to the fact that
the enhanced emission of $\alpha$ and $p$ from the tips of the nucleus
reduces partly the effect of its deformation.  The distribution for the
case when the emission of protons and $\alpha$-- particles is inhibited is
shown on the r.h.s. of Fig. 5 by the short--dashed line. It is seen
that this distribution is very close to the one obtained when the
emission of all three kinds of particles is allowed.
 
\subsection{Decay of $^{126}$Ba}

In the last months a new experiment on fusion and decay of $^{126}$Ba
was performed at the VIVITRON accelerator of the CRN in Strasbourg. The 
DEMON facility was used to detect the outgoing neutrons \cite{Han97}.
Two types of reactions were studied:
\begin{itemize}
\item $^{28}$Si + $^{98}$Mo at $E_{lab}$= 142.*, 165.8, 187.2 and 204 MeV,
\item $^{19}$F + $^{107}$Ag at $E_{lab}$= 128.0 and 147.8 MeV
\end{itemize}
In this way the compound nucleus $^{126}$Ba was produced at four
different excitation energies $E^*$= 84.1, 101.5, 118.5 and 131.7 MeV.
The spin distribution differs in each case since it depends on the energy 
and on the way in which the compound nucleus was produced. 

As the results of these experiments are still analyzed, and neither fusion
cross sections nor multiplicities of outgoing particles are avaible,
we have estimated the fusion cross section using our model described in
\cite{Prz94,Pom94}. The Langevin transport equation was used to describe 
the fusion process. The effect of deformation of the colliding ions was taken 
into account \cite{Pom94}. The results of this numerical simulation are 
presented in Fig. 6, where the differential cross section is plotted in 
form of bins as a function of L for the six different ways in which 
$^{126}$Ba was fused.
\ifig{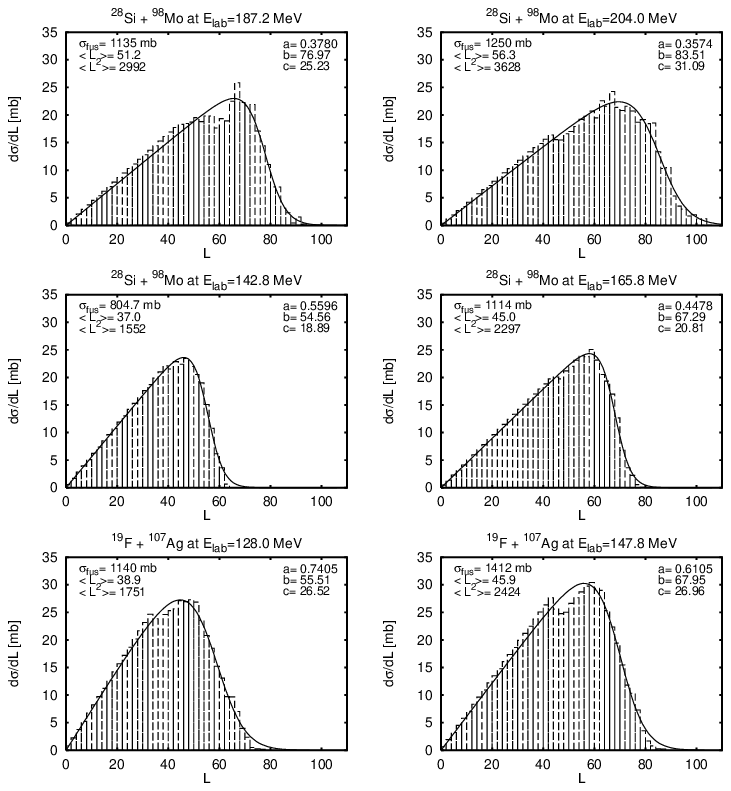}{180mm}{1}{}
%6
Each simulation was performed on basis of 30.000 trajectories. The total
fusion cross section $\sigma_{fus}$ as well as the average angular momentum
and its variance are written in the upper left corner of the plots.
We have also fitted these fusion cross sections by the function (solid line)
$$
{d \sigma\over d L} = {a\cdot L\over 1 + e^{L^2-b^2\over c^2}} \,\,.
$$
The parameters $a, b, c$ characteristic for each distribution are written
in the upper right corners of the plots. We can see in Fig. 6 that the
distribution of the angular momentum $L$ is different in each case.
We predict that the maximal angular momentum of the fused system 
varies from about 60$\hbar$ up to 100$\hbar$ depending on the way in which
$^{126}$Ba is produced.

The initial temperatures of the compound nucleus $^{126}$Ba are rather 
low as the total excitation energy of the system is not very high and
a large fraction of the excitation energy is stored in the form of 
rotational and deformation energy. 
Fig. 7 illustrates how the temperature of $^{126}$Ba changes
as function of the the angular momentum $L$ for given excitation energy.
For each $L$ value the corresponding equilibrium deformation was chosen.
\ifig{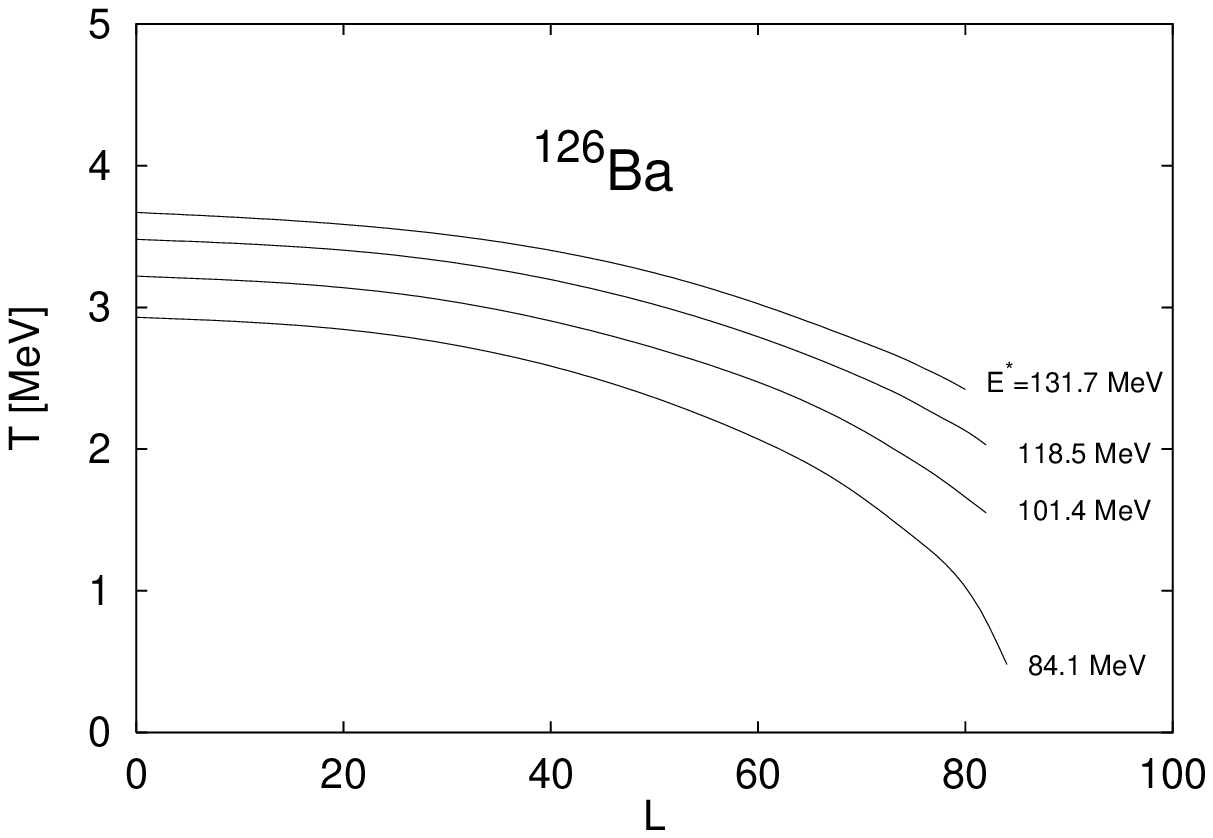}{60mm}{1}{}
%7
It is seen that for the angular momenta $L\geq 80$, which contribute 
mostly to fission of $^{126}$Ba, and the lowest excitation energy 
$E^*$=84.1MeV the initial temperature reaches the value $T\approx$ 0.5 MeV. 
At such a low temperature our model (transport equation for fission 
without superfluidity coupled 
with the Master equation for particle evaporation) is poorly justified.
So one should cosider the estimates of the prefission particles we obtained 
for this energy as only a rough estimate.

Theoretical estimates of the multiplicities of neutron, proton and 
$\alpha$--particles are given in the table:
\vspace{5mm}
\begin{center} 
\begin{tabular}{|c|c|c|c|c|c|c|c|c|} 
\hline 
Reaction & $E_{lab}$ &$E^*$&$M_n$&$M_p$&$M_\alpha$&$S_n$&$S_p$&$S_\alpha$\\
\hline
&MeV&MeV&-&-&-&-&-&-\\
\hline
      & 204.0 & 131.7 & 2.29 & 0.03 & 0.79 & 2.30 & 0.11 & 4.11 \\
$^{28}$Si + $^{98}$Mo &
        187.2 & 118.5 & 1.71 & 0.00 & 0.09 & 2.19 & 0.09 & 3.71 \\
      & 165.8 & 101.4 & 1.83 & 0.00 & 0.04 & 2.13 & 0.05 & 3.20 \\
      & 142.8 &  84.1 & 0.27 & 0.04 & 0.88 & 2.23 & 0.02 & 2.56 \\
\hline
$^{19}$F + $^{107}$Ag &
        147.8 & 118.5 & 1.99 & 0.00 & 0.16 & 2.13 & 0.09 & 3.84 \\
      & 128.0 & 101.5 & 1.80 & 0.01 & 0.06 & 2.11 & 0.05 & 3.30 \\
\hline
\end{tabular}
\end{center}
The thermal excitation energies of $^{126}Ba$ are rather low, so we have 
assumed here the preformation factor for $\alpha$--particles $f_\alpha$=1 
as for cold nuclei.
Just for comparison we have presented also the multiplicities of 
these particles ($S_i,\,\, i=n, p, \alpha$) connected with the evaporation
residua.  One can see from the table that a rather large  number 
($\approx $ 4)
of $\alpha$--particles is emitted by the residua. It is due to the fact that 
the nucleus $^{126}$Ba has a rather small Coulomb barrier for 
of $\alpha$--particles. Furthermore, the large average orbital momentum 
of the mother nucleus favores the emission of $\alpha$--particle. 
Due to these two 
effects even the emission of $\alpha$--particles  with very low kinetic 
energy becomes possible. As a consequence the $\alpha$ emission 
competes significantly with neutron emission and becomes even larger for 
the evaporation residua.

If the particles are emitted in coincidence with fission they are emitted 
at very large deformation ($R_{12}\approx 2$), wheras if they are emitted 
in coincidence with evaporation residua the emission occurs at smaller 
deformation ($R_{12} \approx$ 1.2). 
This effect could be observed in the angular
distribution of emitted neutrons: More particles should be emitted in
the direction perpendicular to the reaction plane as in the reaction
plane. This effect will not be so visible for $\alpha$--particles
because the reduction of the Coulomb barrier is largest at the tips of 
the highly
deformed nucleus and at the same time the collective centrifugal force
acting on the $\alpha$--particle is the largest in this peripheral region 
what favoures  emission in the reaction plane.

\section*{Summary}

Our results demonstrate the influence of the nuclear deformation and
of the collective rotation on 
the evaporation of light particles from excited 
nuclei. The dependence on the deformation plays an important
role also for the competition between fission and particle
emission and might modify the limits which were determined for the 
nuclear friction force \cite{Str91} from the experimental data 
on evaporation and fission. 

Due to the strong dependence
of the fission probability on the initial angular momentum, it
is very important to obtain a precise information on the angular
momentum distribution of the initial ensemble of compound
nuclei. The outcome of the competition between light particle
emission and fission  depends strongly on the initial angular momenta.
The reason is that the rotational angular momentum of the
nucleus has a very noticeable influence on the height of the
fission barrier which decreases as a function of increasing
angular momentum. Thus, at high angular momentum, nuclear
fission can compete more effectively with evaporation.

We hope that the angular distribution of emitted particles, especially
neutrons, depends on the deformation of the source nuclei sufficiently
sensitively so as to determine the deformation from such measurements.
The experimental data on the angular distribution of emitted neutrons,
protons, and $\alpha$--particles from aligned rotating deformed nuclei
would be of great interest for these studies.

\subsection*{Acknowledgment}

Krzysztof Pomorski gratefully acknowledges the warm hospitality extended
to him by the Theoretical Physics Group of the Technische Universit\"at
M\"unchen as well as to the Deutsche Forschungs Gemeinshaft for granting 
a~guest professor position . 
This work is also partly supported by the Polish State Committee for
Scientific Research under Contract No.  2P03B01112.


\begin{thebibliography}{99} 
\bibitem{Str91} E. Strumberger, K. Dietrich,K. Pomorski,
                Nucl. Phys. {\bf A529} (1991) 522.
\bibitem{Pom96} K. Pomorski, J. Bartel, J. Richert, K. Dietrich,
                Nucl. Phys {\bf A 605} (1996) 87.
\bibitem{Wei39} V. Weisskopf, Phys. Rev. {\bf 52} (1937) 295.
\bibitem{Kra40} H. Kramers, Physica {\bf 7} (1940) 284.
\bibitem{Gra79} P. Grang\'e, H.C. Pauli,H.A. Weidenm\"uller,
                Phys. Lett. {\bf B88} (1979) 9; ~Zeit. Phys. {\bf A296}
                (1980) 107.
\bibitem{Fro92} P. Fr\"obrich, Nucl. Phys. {\bf A545}(1992) 87c.
\bibitem{Til92} G.R. Tillack, R. Reif, A. Sch\"ulke, P. Fr\"obrich,
                H.J. Krappe, H.G. Reusch, Phys. Lett. {\bf B296} (1992) 296.
\bibitem{Abe90} Y. Abe, N. Carjan, M. Ohta, T. Wada, {\em Proc. IN2P3--RIKEN
                Symp. on Heavy--Ion Collisions}, Obernai, 1990, France
\bibitem{Hil92} D. Hilscher, H. Rossner, Ann. Phys. Fr. {\bf 17} (1992) 471 
\bibitem{Gon90} M. Gonin, L. Cooke, K. Hagel, Y. Lou, J.B. Natowitz, 
                R.P. Schmitt, S. Shlomo, B. Srivastava, W. Turmel, 
                H. Ustunomiya, R. Wada, G. Nardelli, G. Nebbia, G. Viesti, 
                R. Zanon, B. Fornal, G. Prete, K. Niita, S. Hannuschke, 
                P. Gonthier, B. Wilkins, Phys. Rev. {\bf C42} (1990) 2125.
\bibitem{Han97} F. Hanappe et al., {\em private communication}
\bibitem{Blo86} J. B\l ocki, H. Feldmeier, W.J. Swiatecki, 
                Nucl. Phys. {\bf A459} (1986) 145.
\bibitem{Bla80} M. Blann, Phys. Rev. {\bf C21} (1980) 1770.    
\bibitem{Bar95} J. Bartel, K. Mahboub, J. Richert and K. Pomorski, 
                Zeit. Phys. {\bf A354} (1996) 59.
\bibitem{Die96} K. Dietrich et al, {\em in preparation}.
\bibitem{Prz94} W. Przystupa, K. Pomorski, Nucl. Phys. {\bf A572} (1994) 153.
\bibitem{Pom94} K. Pomorski, W. Przystupa, J. Richert, Acta Phys. Polon.
                {\bf B25} (1994) 751.
\end{thebibliography}
\end{document}